\documentclass[AMA,Times1COL]{WileyNJDv5}

\articletype{Research Article}%

\received{Date Month Year}
\revised{Date Month Year}
\accepted{Date Month Year}
\journal{Journal}
\volume{00}
\copyyear{2024}
\startpage{1}

\begin{document}

\title{Identifying Influential nodes in Brain Networks via Self-Supervised Graph-Transformer}

\author[1]{Yanqing Kang}
\author[1]{Di Zhu}
\author[1]{Haiyang Zhang}
\author[1]{Enze Shi}
\author[1]{Sigang Yu}
\author[1]{Jinru Wu}
\author[1]{Xuhui Wang}
\author[1]{Xuan Liu}
\author[2]{Geng Chen}
\author[3]{Xi Jiang}
\author[4]{Tuo Zhang}
\author[1]{Shu Zhang}

\authormark{Yanqing Kang \textsc{et al.}}
\titlemark{Identifying Influential nodes in Brain Networks via Self-Supervised Graph-Transformer}

\address[1]{\orgdiv{Center for Brain and Brain-Inspired Computing Research, School of Computer Science}, \orgname{Northwestern Polytechnical University}, \orgaddress{\state{Xi'an}, \country{China}}}

\address[2]{\orgdiv{School of Computer Science}, \orgname{Northwestern Polytechnical University}, \orgaddress{\state{Xi'an}, \country{China}}}

\address[3]{\orgdiv{School of Life Science and Technology}, \orgname{University of Electronic Science and Technology of China}, \orgaddress{\state{Chengdu}, \country{China}}}

\address[4]{\orgdiv{School of Automation}, \orgname{Northwestern Polytechnical University}, \orgaddress{\state{Xi'an}, \country{China}}}

\corres{Shu Zhang, Center for Brain and Brain-Inspired Computing Research, School of Computer Science, Youyi campus of Northwestern Polytechnical University ADD: 127 West Youyi Road, Beilin District, Xi'an Shaanxi, 710072, P.R.China.\email{shu.zhang@nwpu.edu.cn}}

\presentaddress{Youyi campus of Northwestern Polytechnical University ADD: 127 West Youyi Road, Beilin District, Xi'an Shaanxi, 710072, P.R.China.}


\abstract[Abstract]{Studying influential nodes (I-nodes) in brain networks is of great significance in the field of brain imaging. Most existing studies consider brain connectivity hubs as I-nodes such as the regions of high centrality or rich-club organization. However, this approach relies heavily on prior knowledge from graph theory, which may overlook the intrinsic characteristics of the brain network, especially when its architecture is not fully understood.  In contrast, self-supervised deep learning dispenses with manual features, allowing it to learn meaningful representations directly from the data. This approach enables the exploration of I-nodes for brain networks, which is also lacking in current studies.  This paper proposes a Self-Supervised Graph Reconstruction framework based on Graph-Transformer (SSGR-GT) to identify I-nodes, which has three main characteristics. First, as a self-supervised model, SSGR-GT extracts the importance of brain nodes to the reconstruction. Second, SSGR-GT uses Graph-Transformer, which is well-suited for extracting features from brain graphs, combining both local and global characteristics. Third, multimodal analysis of I-nodes uses graph-based fusion technology, combining functional and structural brain information. The I-nodes we obtained are distributed in critical areas such as the superior frontal lobe, lateral parietal lobe, and lateral occipital lobe, with a total of 56 identified across different experiments. These I-nodes are involved in more brain networks than other regions, have longer fiber connections, and occupy more central positions in structural connectivity. They also exhibit strong connectivity and high node efficiency in both functional and structural networks. Furthermore, there is a significant overlap between the I-nodes and both the structural and functional rich-club.  Experimental results verify the effectiveness of the proposed method, and I-nodes are obtained and discussed. These findings enhance our understanding of the I-nodes within the brain network, and provide new insights for future research in further understanding the brain working mechanisms.}



\maketitle



\section{INTRODUCTION}\label{sec1}

The human brain is an extremely complex system that coordinates human cognition, behavior, etc.\cite{barbey2018network,bressler2010large,park2013structural}. It involves a lot of connections between neurons. Studies have shown for a long time that the brain can be represented as a network, with different regions of the brain connected to each other in specific patterns\cite{bullmore2011brain}. It is generally believed that there is a group of influential regions in the brain network that play an important role. For example, the highly connected regions in the brain network are called brain network hubs. These hub regions are highly connected to the rest of the brain and take a central position in neural information processing, facilitating communication between different parts of the brain\cite{van2013network}. Understanding the functions of these influential regions is crucial for unraveling the intricate workings of the brain, shedding light on mechanisms underlying cognition, learning, and memory. Moreover, the identification and study of these regions hold significant implications for the investigation of neurological disorders, as disruptions in these influential regions may contribute to the manifestation of conditions such as schizophrenia, depression, and other cognitive impairments.

Over the past few decades, researchers have come up with various ways to detect influential regions of the brain, such as the brain hubs. Early on, the hubs of the structural and functional networks of the brain were studied, separately\cite{bullmore2009complex,van2011rich,power2011functional,zhu2013dicccol,liao2017small,betzel2017multi,kim2020rich,zhang2020cortical}. For example, in terms of structural networks, Gong et al. used Diffusion Tensor Imaging (DTI) to construct brain structural networks and analyzed their topological properties. It is discovered that the network has the small-world property and the nodes with high centrality are brain hubs \cite{gong2009mapping}. Heuvel et al. analyzed the relationship between the hubs of the structural network and proved that the brain hubs formed the rich-club and played an important role in the information transmission and integration of the brain\cite{van2011rich}. In terms of functional networks, Power et al. adopted subgraph detection algorithm for functional networks and discovered that there are highly interconnected regions among subgraphs, and these regions are considered as key hubs for functional integration\cite{power2011functional}. Since it is difficult for single modal approaches to provide a comprehensive understanding of the working mechanism of the brain, some multimodal studies have combined structure and function to analyze brain hubs\cite{sui2012review,grayson2014structural}. 

However, whether from the perspective of single mode or multimodal, most studies on the detection of brain hubs are based on prior knowledge or artificial design. For example, some of the above studies, from the point of view of graph theory attributes, believe that the regions with high centrality or make up the rich-club organization are the hubs. But the true architecture of brain networks is not fully understood, and relying excessively on prior knowledge may lead to overlooking the intrinsic characteristics of the brain network itself. So, it is eagerly necessary to explore the influential regions from a data-driven perspective rather than manual features. In addition, deep learning has been increasingly applied to brain neuroimaging field\cite{zhao2021deep}. For example, the graph convolutional network (GNN) and convolutional neural network (CNN) are used to classify diseases and obtain corresponding biomarkers \cite{kawahara2017brainnetcnn,wang2019functional,jiang2020hi,wang2019spatial,bi2020functional,li2021braingnn,bi2023boosting,bi2023structure}. But these supervised approaches cannot be applied to detect influential regions of brain networks because they require labels to achieve the classification related tasks. So, it is also crucial to develop the self-supervised deep learning approach to automatically identify the influential regions, which are currently lacking in research. Encouraged by the urgent needs above, we try to combine the self-supervised \cite{qiao2018data,liu2021self,liu2022graph} method in deep learning with the influential regions detection of the brain. Moreover, Self-supervised learning has been proven to detect meaningful features from brain data \cite{millet2022toward,thomas2022self}. Wen et al. applied self-supervised learning with mask autoencoder on brain network analysis, reporting significant improvements in disease classification task \cite{wen2023graph}. Weis et al. proposed the self-supervised method based on contrast learning and transformer for graph representation learning, successfully applying it to the classification of excitatory neuron morphologies \cite{weis2021self}. Based on the above findings, it is a novel and feasible attempt to use graph self-supervised learning to detect the influence regions of brain networks. 

In this paper, we propose a self-supervised graph reconstruct framework based on Graph-Transformer (SSGR-GT) to identify influential nodes (I-nodes) of brain networks. We represent the brain as a graph, where the nodes of the graph are the regions of interest (ROIs, and in this paper, ROI and node have the same meaning) defined by the brain atlas, and the edge information of the graph comes from the structural connectivity of the brain ROIs, and the node features of the graph come from the functional connectivity of the brain ROIs. A self-supervised model is designed to reconstruct the input brain graph and extract nodes that are important to the reconstruction task. We think that these nodes, which are important for the reconstruction task, are the I-nodes of brain networks. There are two reasons: first, its node features contain rich functional information required for the graph reconstruction. This suggests that these nodes are important in the brain functional connected network. They may be involved in more functional networks and play a key role in the functional integration of the brain. Second, these nodes have strong structural connections with other nodes and are located at the center of the entire graph. This means that these nodes have an important influence on the stability of the brain structural network. The above characteristics mean that these nodes play an important role in the structural connectivity and functional transmission of brain, which can be regarded as the I-nodes of brain networks.

Note that this study is considerably extended from our preliminary work \cite{kang2023exploring}. For methodological contribution, we develop and apply a new mask mechanism to improve the self-supervised model used, and conduct more comprehensive experiments for analysis and interpretation, such as model stability and result comparison. Our approach has three major characteristics. First, a self-supervised learning model is designed to detect the I-nodes of brain networks by analyzing data in a completely data-driven manner. Specifically, we designed a self-supervised model with the task of reconstructing the input brain graph and extracted the importance of nodes to the reconstruction task. Nodes that are important to the reconstruction task are considered to be I-nodes. Second, the self-supervised model we designed is based on Graph-Transformer, which is very suitable for feature extraction of brain graph. It consists of GNN and Transformer \cite{vaswani2017attention}. Graph convolutional network (GCN)\cite{zhang2019graph}  is used to establish a close relationship between the neighbor nodes and make the model focus on local information. Transformer applies a self-attention mechanism \cite{vaswani2017attention} to all nodes so that the model can focus on global information at the same time. Third, multimodal analysis of I-nodes is used on the brain graph. It maps the structural and functional information of the brain to the edges and nodes of the graph, respectively. And deep learning is used to fuse multimodal information to provide more complete understanding of brain connectivity and functional organization.

\section{METHOD}\label{sec2}

\subsection{Overview}

\begin{figure*}[htbp]
\centerline{\includegraphics[width=1\linewidth]{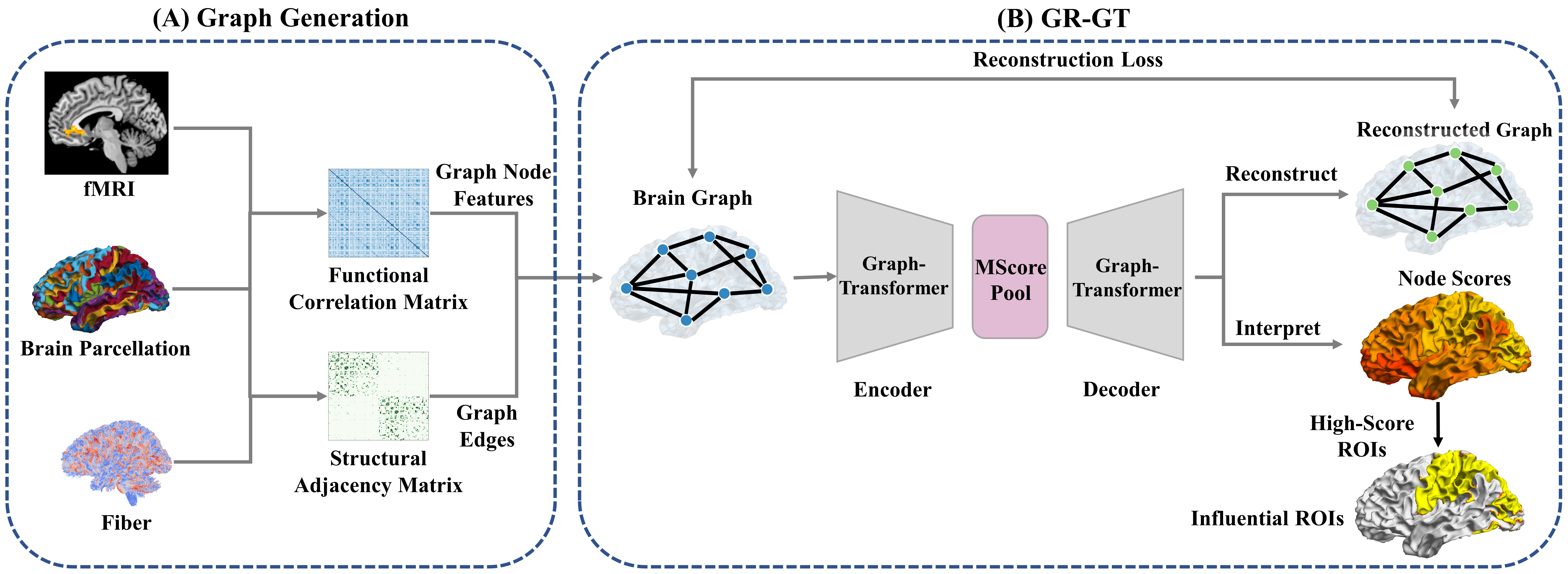}}
\caption{The pipeline of the proposed SSGR-GT framework. (A) represents the graph generation of brain. (B) represents the GR-GT module used to reconstruct the graph and obtain node scores. The module consists of three parts, including the Encoder module, MScore Pool, and Decoder module. \label{fig1}}
\end{figure*}

The pipeline of the SSGR-GT framework proposed in this paper is shown in \textbf{Fig. \ref{fig1}}. The method analyzes I-nodes with the joint analysis of brain functional and structural profiles. The brain is represented as a graph, and through the self-supervised model we can ultimately obtain a reconstructed graph and contribution scores of nodes. Nodes with high scores are more important for reconstruction tasks and are considered as I-nodes of brain networks. 

The framework consists of graph generation and Graph Reconstruction based on Graph-Transformer (GR-GT) module. In the graph generation part, we initialize the brain surface into 148 ROIs based on Destrieux Atlas\cite{destrieux2010automatic}  and use them as nodes of the graph. Some studies have integrated structural and functional brain information in a graph representation \cite{dsouza2021m,zhang2022differentiate}. Inspired by their approach, we use functional and structural information of the brain to generate node features and edge features of the graph, respectively. The GR-GT module consists of three main components, namely Encoder, MScore Pool and Decoder. The Encoder module is used to extract the node representation of the brain graph, and the Decoder module is used to reconstruct the nodes’ features of the brain graph. To extract the contribution scores of nodes, we design the MScore Pool, which comes after the Encoder module. In order to improve the learning ability and stability of the model, we add a mask mechanism to the MScore Pool. The node representations of the low-score nodes are replaced by the mask embedding, which together with the node representations of the high-score nodes are used as the input to the Decoder module. Finally, we extract the scores of nodes from the MScore Pool, and the high-score nodes are considered as I-nodes.

\subsection{Data and Preprocessing}
We use the HCP 900 dataset \cite{van2012human} and randomly select 98 subjects from it. The participants in the HCP 900 dataset consist of healthy adults with ages ranging from 22 to 36 years old. Among the 98 subjects selected, 55 are female and 43 are male. T1-weighted magnetic resonance imaging (MRI) data is used to reconstruct the brain cortical surface, diffusion-weighted MRI (dMRI) is utilized to reconstruct the fiber bundles from white matter, functional MRI (fMRI), including resting state fMRI (rs-fMRI) and task fMRI, shows the functional changes of the brain. Among them, task fMRI contains a total of seven tasks, which are EMOTION, GAMBLING, LANGUAGE, MOTOR, RELATION, SOCIAL, and Working Memory (WM). Briefly, the EMOTION cognition task involves making judgments about facial expressions, including negative (fearful and angry) and neutral faces. The GAMBLING reward task require participants to play a card-guessing game to win or lose money by guessing the numbers on the cards. The LANGUAGE task consists of language condition related to listening to stories and math condition related to arithmetic questions. The MOTOR tasks require subjects to perform motor tasks, such as tap their left/right fingers, squeezing their left/right toes, or moving their tongue. The RELATIONAL task asks subjects to judge whether two sets of objects agree on the same dimension (e.g., shape or texture). In the SOCIAL cognition task, participants view a video clip where shapes interact, then determine whether the interaction suggests social behavior. Last, the WM task is a variant of the N-back task. Please refer to Barch 
et al for more details \cite{barch2013function}. 

Standard Freesurfer pipeline including tissue-segmentation and white matter surface (inner surface) reconstruction \cite{fischl2002whole} is proposed to preprocess the T1-weighted MRI. For fMRI, we adopted the graycoordinate system \cite{glasser2013minimal} as the platform to extract the fMRI time sequence for each surface vertex. For dMRI, we followed the method in Van den Heuvel et al. \cite{van2012high} to use deterministic tractography to derive white matter fibers, and reconstructed $ 5\times10^4$ fiber tracts for each subject.

To jointly use these three modalities, we aligned them into the same space. A linear image registration method (FLIRT) \cite{jenkinson2002improved} and a nonlinear one (FNIRT) \cite{jenkinson2012fsl} were cascaded to transform and warp T1-weighted MRI to the dMRI space. The preprocessed fMRI uses graycoordinate system, which is located at the same space as T1-weighted MRI, and the vertex-wise correspondence between them can be directly established.
\subsection{Self-supervised Graph Reconstruct Framework based on Graph-Transformer (SSGR-GT Framework) }
\subsubsection{Graph Generation}
We define the brain as an undirected graph $G=\{V,E,X\}.$ $V=\{v_i|i\in 1,2,
\ldots,N\}$represents the set of N nodes of the graph. $X\in R^{N\times D}$ represents the feature matrix of graph nodes, where $D$ denotes the length of the node feature.  $E\in R^{N\times N}$ represents the adjacency matrix of the graph.

\begin{enumerate}[1)]
\item \textbf{Graph Nodes with Generated Features} 

The brain surface is divided into 148 ROIs and used as nodes of the graph. fMRI signals are selected to express node features. First, we obtain the fMRI signals of each ROI by averaging the fMRI signals of all vertices within the ROI. Then, the Pearson correlation coefficient of the signals between each pair of ROI is calculated to obtain the functional similarity matrix. The functional similarity matrix is used as the feature matrix  of X the graph nodes. The length of each node feature vector D is 148.
\item \textbf{Graph Edge}

For each pair of ROIs, we calculate the total number of fibers directly connected to them and then divide it by the geometric mean of their areas, thus obtaining the structural connectivity matrix S . We set the threshold $ t_s $ for S to make it sparse and do the binarization to obtain the adjacency matrix E of the graph. $ E_{i,j} =1$ means that nodes i and j are connected, otherwise $ E_{i,j}=0$ .

Finally, we obtained $ 98\times8=784$ brain graphs. 98 represents the number of individuals, and 8 represents eight fMRI data. For different individuals, the adjacency matrices between the brain graphs are different. For the same individual, the adjacency matrices between the brain graphs are the same and the node features are different.

\end{enumerate}

\subsubsection{GR-GT Module}
\begin{figure*}[htbp]
\centerline{\includegraphics[width=0.8\linewidth]{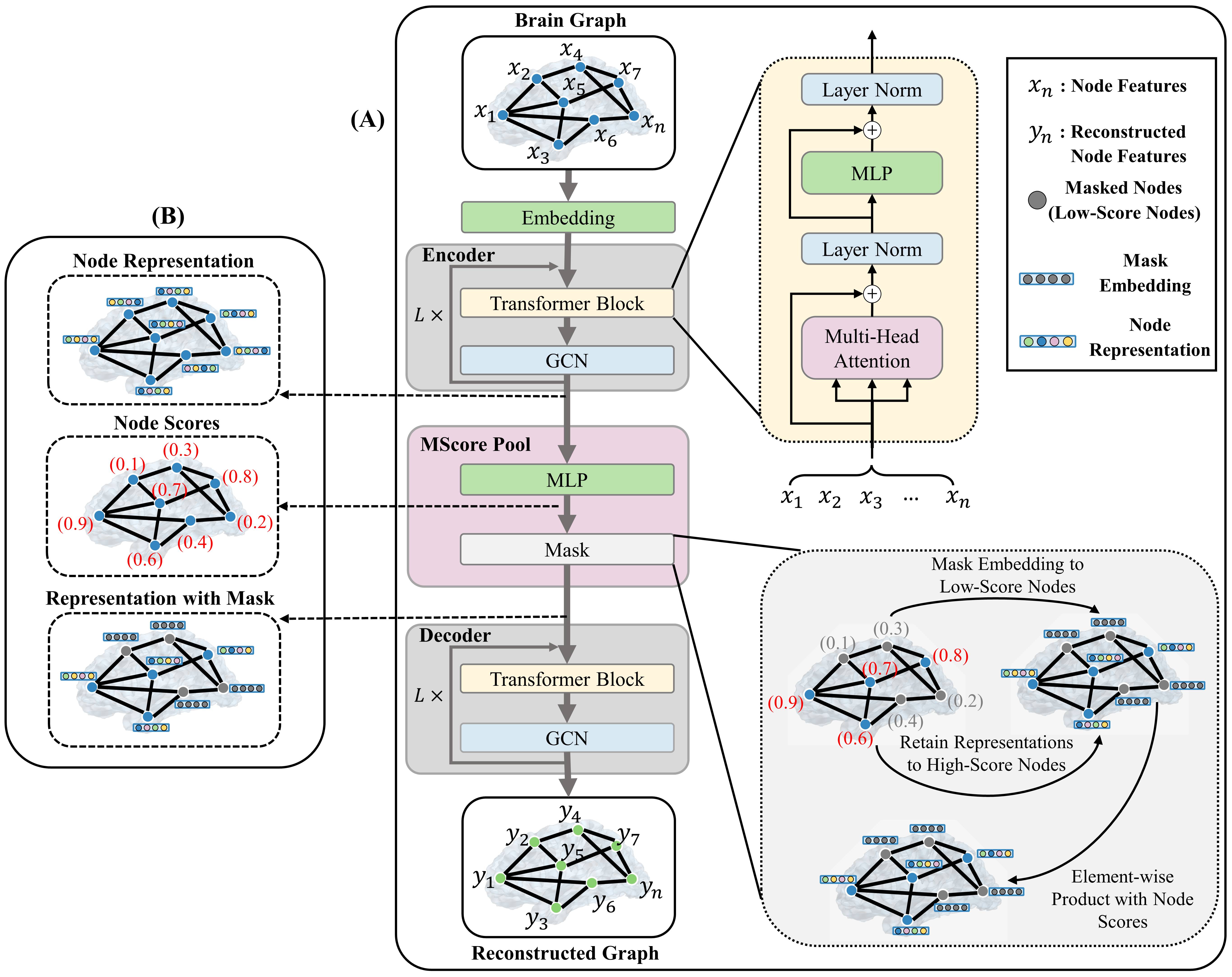}}
\caption{The detail of the GR-GT module. (A) represents the network architecture of the module. (B) represents the data representation in the module process. \label{fig2}}
\end{figure*}

In the GR-GT module, we adopt the Encoder-Decoder structure to implement the graph reconstruction task, as shown in \textbf{Fig. \ref{fig2}}. The Encoder module is applied to extract node representations of the graph, then the MScore Pool is used to obtain the contribution scores of nodes, and finally the Decoder module is used to reconstruct the node features of the graph.
\begin{enumerate}[1)]
\item \textbf{Encoder Module}

The Encoder module is applied to extract node representations of the graph, as shown in \textbf{Fig. \ref{fig2} (A)}. And the Encoder module in this paper is based on UGformer \cite{nguyen2022universal}. It combines GCN, which establishes a close relationship between neighbor nodes, and Transformer, which applies a self-attention mechanism to all nodes so that the model can focus on both local and global information. For each layer of the Encoder module, it first adopts the Transformer network and applies the self-attention mechanism to all nodes of the graph instead of within neighboring nodes. Since all nodes in the self-attentive layer interact with each other to form a complete network, the structural information of the graph is ignored. To overcome this defect, the GCNs is used to utilize the structural information of the graph after the Transformer network. The process is shown as Eq. (1) and Eq. (2)

\begin{equation}\label{eqn-1}
      H^{'(k+1)}=Attention_v(H^{(k)}Q^{(k)},H^{(k)}K^{(k)},H^{(k)}V^{(k)})
\end{equation}

\begin{equation}
    H^{(k+1)}=GCN(E,H^{'(k+1)})
\end{equation}

\noindent
where $H^{(k)}\in R^{N*d}$ is the nodes’ representations of the graph at the -th layer of the Encoder module, $H^{(0)}=X;Q^{(k)}\in R^{d*d},K^{(k)}\in R^{d*d},V^{(k)}\in R^{d*d}$are query-projection, key-projection and value-projection weight matrix; $V$ is the set of all nodes of the graph, and  is the adjacency matrix. The detailed expression for Eq. (1) is as follows:
\begin{equation}
\begin{aligned}
      u_v^{(k)}=Lnorm(h_v^{(k)}+Att(h_v^{(k)}))\\
    h_v^{(k+1)}=Lnorm(u_v^{(k)}+Trans(u_v^{(k)}))
\end{aligned}
\end{equation}
where $h_v^{(k)}$ is the vector of node $v$, $Trans$ and $Att$ denote a MLP layer and a self-attention layer respectively, and $Lnorm$ denotes normalization. The detailed expression for $Att$ is as follows:
\begin{equation}
    \begin{aligned}
        Att(h_v^{(k)})=\sum\nolimits_{v^{'} \in V}{\alpha _{v,v^{'}}^{(k)}(V^{(k)}h_{v^{'}}^{(k)})}\\
        \alpha _{v,v{'}}^{(k)}=softmax(\frac{(Q^{(k)}h_v^{(k)})\cdot((K^{(k)}h_{v^{'}}^{(k)}))}{\sqrt{d}})
    \end{aligned}
\end{equation}
where $ \alpha _{v,v{'}}^{(k)}$ is an attention weight between nodes $v$ and $v^{'}$. The detail for Eq. (2) is as follows:
\begin{equation}
    \begin{aligned}
        H^{(k+1)}=\sigma({\hat{E}}H^{'(k+1)}W^{(k)} )\\
        \hat{E}=D^{-\frac{1}{2}}(E+1)D^{-\frac{1}{2}}\\
        D_{ii}=\sum_{j}E_{ij}+1
    \end{aligned}
\end{equation}
where $W^{(k)}$ is the weight matrix, $I$ is the identity matrix, and $\sigma$ is the nonlinearity function.

\item \textbf{MScore Pool}

Inspired by the pooling operation of the Graph U-Nets \cite{gao2019graph}, we design the MScore Pool to get the contribution score of each node in the graph for the reconstruction task. In the reconstruction task, in order to improve the learning ability of the model, the data is usually processed with noise or mask. Similar to the mask mechanism in MAE \cite{he2022masked}, we add a mask mechanism to MScore Pool. Low-score nodes are masked according to a certain mask ratio, and their node representations are replaced by the mask embedding that indicates the node features to be reconstructed. The mask embedding and the unmasked node representations are used together as inputs to the Decoder. The output of the Encoder module is the graph nodes’ representations$Z \in R^{N*d}$, which is used as the input of the MScore Pool. After an MLP layer and softmax, $Z$ is mapped to $p \in R^{N*1} $ . $p$is seen as the contribution scores of the graph nodes. According to a certain ratio $k$, the nodes with low scores are masked. That is to use mask embedding, a shared learning vector $q$, to represent masked nodes. And the representations of nodes with high scores is retained. We concatenate the retained nodes and the masked nodes in the original order, and do the element-wise product operation with the score $p$ as the input to the Decoder module. It is calculated as Eq. (6). 
\begin{equation}
    \begin{aligned}
        p=softmax(MLP(Z))\\
        idx=TopK(p,1-k)\\
        \widehat{Z} = cat(Z(idx,:),q)\\
         Z^{'}= \widehat{Z}\odot p
    \end{aligned}
\end{equation}
where $k$ represents the ratio of masked nodes, $idx$ represents the indexes corresponding to the first $1-k$ maximum elements in $p$, $q$ denotes the embedding vector of the masked nodes, $cat$ denotes concatenating the node representation with the masked embedding in the original order, $\odot$ denotes the element-wise product, and $ Z^{'}$ denotes the input of the Decoder module.
\item \textbf{Decoder Module}

The input to the Decoder module is the graph node representations with mask embedding, and the output is reconstructed node features of the original brain graph. The Decoder module has the same architecture as the Encoder module, both of which are based on the UGformer module. Our GR-GT module reconstructs the graph by predicting the node features of the graph, and the loss function is mean square error (MSE) between the feature matrix $Y$ of reconstructed graph nodesand feature matrix $X$ of input graph nodes. 
\end{enumerate}

\subsection{Ranking ROIs via MScore Pool }
Our goal is to design a self-supervised graph reconstruction model and get the contribution score of each ROI to graph reconstruction. In Section 2.3, we design the MScore Pool to extract the contribution scores of the ROIs. Since the model provides scores of ROIs for each of individual, and what we need is scores of ROI at the group-level, in Section 3.2, we conduct statistical analysis on the scores of all individuals to obtain scores of ROI at the group-level. After getting the group-level scores, we divide the brain ROIs into three different scales according to the score rankings. ROIs in Scale-1, with the higher scores, are considered as I-nodes of brain networks. In the following sections, we will verify the rationality of the selected ROIs through experiments, that is, the ROIs with high scores are the I-nodes. 
\subsection{Evaluating the Hub Attributes of the High-Score ROIs}
As shown in Section 2.4, we treat Scale-1 ROIs as I-nodes. In Section 3.3 and 3.4, we conduct experiments to see if Scale-1 ROIs are I-nodes in the brain. On the one hand, we analyze the characteristics of the brain functional network and structural connectivity for ROIs. For brain function, studies believe that the brain performs specific functions in the form of functional networks \cite{raichle2009brief}. Therefore, we use dictionary learning to decompose and acquire multiple functional networks \cite{lv2014holistic} to explore the participation of ROIs in functional networks. For brain structure, the way that the fibers are connected is very important to it, which is responsible for communication in the brain. Therefore, we explore fiber connection patterns for ROIs, such as fiber length. On the other hand, as the application of graph theory analysis in brain networks has been very mature and extensive  \cite{bullmore2009complex}, here we also adopt the graph theory method to verify the selected I-nodes. First, we construct the functional and structural connectivity networks of the brain, which are obtained by some processing of the functional correlation matrix and the structural connectivity matrix in Section 2.3. Then, we use node centrality and network efficiency metrics to measure the importance of ROIs. There are three metrics of centrality, namely, degree centrality, closeness centrality, and PageRank centrality. And we use global efficiency to represent network efficiency. ROIs with high metrics generally correspond to important ROIs of the brain.
\subsection{Analyzing Foundational Elements for Coordinated Structure and Multi-tasks}
In Section 3.1, we treat the fusion of structure and each kind of brain functional MRI data as a separate experiment, and different experiments obtain different I-nodes. In Section 3.5, we conduct a joint analysis of the I-nodes in all the experiments and hypothesize that some I-nodes and connections remain common across multiple experiments, coordinating the fundamental architecture of the brain with its various functions. These common I-nodes and connections likely support the fundamental integrative capacities that shape higher cognition. Identifying and characterizing these I-nodes and connections that remain coordinated amid changing functions provides key insight into how brain structure and function are linked in a way that maintains stability.
\subsection{Evaluating Reproducibility of I-nodes on New Brain Atlas}
This paper proposes a framework to explore the scores of different brain ROIs and to get the I-nodes. Therefore, the stability and repeatability of model results are very important. In Section 3.6, we use the new atlas to verify the stability of the results. Specifically, we use a new partitioned atlas ($Atlas\_2$) \cite{glasser2016multi} on the same data set and regenerate the brain graphs for the experiment. The $Atlas\_2$ is based on functional-structural multimodal partitioning, which divides each hemisphere into 180 ROIs (removing the corpus callosum), which is more refined than the partition of $Atlas\_1$ (Destrieux Atlas). The experimental results of the two atlases are compared, and the stability of the experiment is verified by the overlap of I-nodes.

\section{EXPERIMENTS AND RESULTS}\label{sec3}

\subsection{Experimental Setup and Performance}
\subsubsection{Experimental Setup}
A total of 98 individuals is evaluated by the proposed method. Each of individual has eight brain graphs, which have the same topology corresponding to the structural information and different node features corresponding to eight kinds of functional information. We set up eight sets of experiments, each of which is trained independently. Each set of experiments use one of the brain graphs from 98 individuals, which corresponded to the same kind of functional information. These eight sets of experiments are called Resting State-Structure (RS), EMOTION-Structure (E), GAMBLING-Structure (G), LANGUAGE-Structure (L), MOTOR-Structure (M), RELATION-Structure (R), SOCIAL-Structure (S), and WM-Structure (W). For the Encoder and Decoder modules, we set the number of layers (L) to 1 and the number of heads in self-attention to 4. The model is trained for 200 epochs using the Adam optimizer to update parameters. The mask ratio k of nodes is 0.5. 
\subsubsection{Model Performance}
\begin{figure*}[htbp]
\centerline{\includegraphics[width=0.8\linewidth]{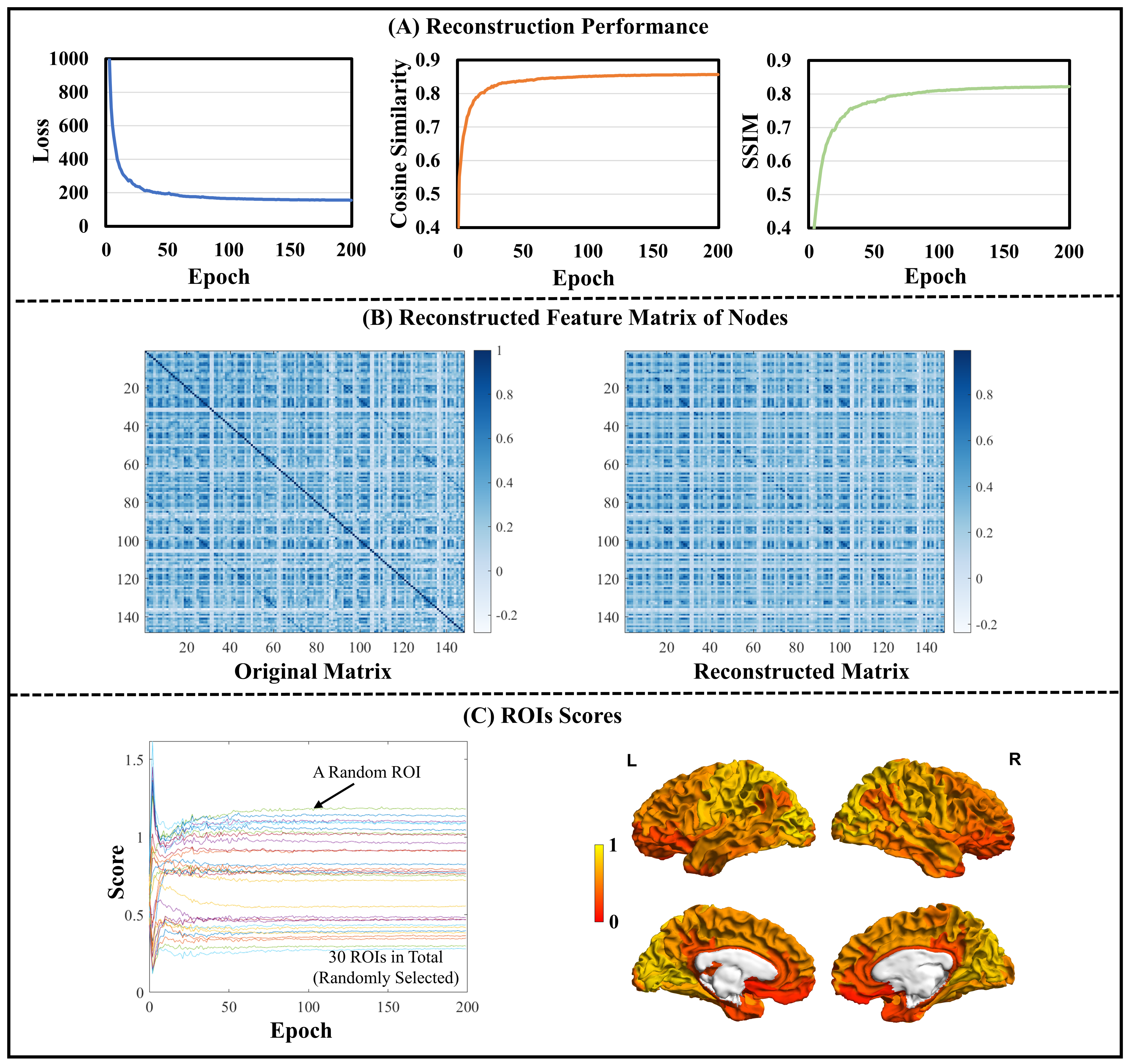}}
\caption{Model Performance for the RS. (A) Model reconstruction quality. (B) Comparison of the original and reconstructed matrices for a randomly selected individual. (C) Changes in scores for selected ROIs and their visualization across brain regions.\label{fig3}}
\end{figure*}

Taking the RS experiment as an example, we show the model performance in \textbf{Fig. \ref{fig3}}.

\begin{enumerate}[1)]
    \item \textbf{Effect of Reconstruction}
    \hangafter 0
    \hangindent 0em

On the one hand, the loss value of the experiment decreases continuously with epoch and eventually becomes relatively stable. On the other hand, to verify the effect of the reconstruction graph, we calculate the cosine similarity and Structural Similarity Index Measure (SSIM) \cite{wang2004image} between the original feature matrix and the reconstruction feature matrix of graph nodes. While cosine similarity is commonly used, we also innovatively adopt SSIM to evaluate the effect of feature reconstruction. SSIM is designed to evaluate the similarity between two images based on three components: luminance, contrast, and structure. Here, we propose that SSIM could capture the topological structure of the original and reconstruction feature matrices, thus measuring the similarity of their functional connections, making SSIM a reliable similarity metric. \textbf{Fig. \ref{fig3} (A)} shows how cosine similarity and SSIM’s mean over all individuals changes with the epoch, they keep increasing with epoch and finally reach 0.86 and 0.82, respectively.
    
    To the best of our knowledge, an SSIM value of 0.82 represents good reconstruction quality, indicating that SSGR-GT implements the task of reconstructing the graph. Firstly, in the field of computer vision, an SSIM value of 0.82 is considered to indicate high-quality image reconstruction. For instance, SRCNN \cite{dong2015image} achieved an SSIM of 0.8215 on the Set14 dataset for super-resolution tasks with a scaling factor of 3. In addition, to further validate the reconstruction effectiveness, we compare the SSIM-based reconstruction performance of proposed SSGR-GT with GraphMAE \cite{hou2022graphmae}, which stands as a state-of-the-art approach in generative self-supervised learning on graphs. As shown in \textbf{Table \ref{tab1}}, it is evident that our SSGR-GT consistently achieves better SSIM scores across all experiments under mask ratio k="0.5" .  
    
    Besides, we randomly select an individual and visualize its original feature matrix and the reconstruction feature matrix, as shown in \textbf{Fig. \ref{fig3} (B)}. And the cosine similarity and SSIM on this individual are 0.88 and 0.85, respectively.\\
    \begin{table}[!ht]%
    \begin{center}
    \caption{Comparison of GraphMAE and our SSGR-GT reconstruction performance SSIM at $k=0.5$.\label{tab1}}%
    \begin{tabular*}{\textwidth}{@{\extracolsep\fill}ccc@{\extracolsep\fill}}%
    \toprule
    \textbf{Experiments} & \textbf{GraphMAE} & \textbf{SSGR-GT}  \\
    \midrule
    RS&0.53&\textbf{0.82}\\
    E&0.42&\textbf{0.73}\\
    G&0.41&\textbf{0.74}\\
    L&0.32&\textbf{0.73}\\
    M&0.33&\textbf{0.75}\\
    R&0.47&\textbf{0.77}\\
    S&0.44&\textbf{0.76}\\
    W&0.41&\textbf{0.75}\\
    \bottomrule
    \end{tabular*}
    \end{center}
    \end{table}
    \item \textbf{Effect of ROIs’ Score}
    
    With the MScore Pool designed by the model, we obtain the contribution scores of brain ROIs. We randomly select 30 ROIs of 148 ROIs to show the results of their scores change with epoch, as shown in \textbf{Fig. \ref{fig3} (C)}. It shows that the scores of ROIs fluctuate more when the epoch is small and gradually stabilize when the epoch is large, and eventually, the scores can be clearly distinguished between ROIs. And scores of all ROIs are visualized in different shades of color on the brain. 
    \item \textbf{Reconstruction Effect under Different Mask Ratio \textit{k}}

    To explore the reconstruction effect of the model under different mask ratios, we add two ratio values $k=0.25,0.75$  for comparison with $k=0.5$ . The effects of all eight sets of experiments under different mask ratios are shown in \textbf{Table \ref{tab2}}. As can be seen from it, both cosine similarity and SSIM in almost all sets of experiments decrease with increasing ratio, indicating that the fewer nodes used to reconstruct the graph, the worse the reconstructed effect. 
    
    At the same time, we also do a comparative experiment with $k=0$, that is, without using the mask mechanism. Compared with experiments using the mask mechanism, it has the best reconstruction effect, but the contribution scores become unstable with the increase of the epoch. On the contrary, \textbf{Fig. 3 (C)} shows that the stability of contribution scores. 
    
For the selection of parameter $k$, we evaluate it from the reconstruction effect and ROIs’ score stability. First, taking S experiment as an example, an individual is randomly selected to visualize the reconstruction matrix under $k=0.5$  and 0.75, as shown in \textbf{Fig. \ref{fig4} (A)}. Compared with the original matrix, the cosine similarity and SSIM of $k=0.5$  are 0.85 and 0.79 respectively, and the cosine similarity and SSIM of $k=0.75$  are 0.82 and 0.68 respectively. The reconstruction matrix with $k=0.75$  loses many details and cannot achieve a good reconstruction effect, while the reconstruction matrix with $k=0.5$  retains most of the details. Then, taking experiment RS as an example, we visualized the ROIs’ scores under $k=0.5$ , 0.25, and 0, as shown in \textbf{Fig. \ref{fig4} (B)}. At $k=0.5$ , ROIs’ scores tend to stabilize and show significant differences at smaller epochs. When $k=0.25$  and epoch = 200, the ROIs’ scores still do not converge completely, and there is a trend of rising or falling. When $k=0$ , that is, when no masking mechanism is used, there is no convergence trend at all when epoch = 200. With the increase of epoch, the difference of scores between ROIs first increases and then decreases, and there is no stable trend. Considering the reconstruction effect and stability comprehensively, $k=0.5$  is selected as the final parameter of the model.
    
    \begin{center}
    \begin{table*}[!ht]%
    \caption{Reconstruction effect under different mask ratio $k$.\label{tab2}}
    \begin{tabular*}{\textwidth}{@{\extracolsep\fill}ccccccc}
    \toprule
   \multicolumn{1}{c}{\multirow{2}{*}{\textbf{Experiments}}} & \multicolumn{3}{c}{\textbf{Cosine Similarity}} & \multicolumn{3}{c}{\textbf{SSIM}} \\\cmidrule{2-7}
    &\textbf{$k=0.25$}  & \textbf{\textbf{$k=0.5$}}  & {\textbf{\textbf{$k=0.75$}}}  & \textbf{\textbf{$k=0.25$}} &\textbf{$k=0.5$}  &\textbf{$k=0.75$}\\
    \midrule
    RS&	0.90&	0.86&	0.83&	0.91&	0.82&	0.73   \\
    E&	0.90&	0.85&	0.82&	0.86&	0.73&	0.60\\
    G&	0.90&	0.85&	0.81&	0.86&	0.74&	0.61\\
    L&	0.91&	0.86&	0.82&	0.86&	0.73&	0.60\\
    M&	0.91&	0.87&	0.90&	0.87&	0.75&	0.67\\
    R&	0.91&	0.86&	0.89&	0.88&	0.77&	0.70\\
    S&	0.92&	0.87&	0.85&	0.87&	0.76&	0.64\\
    W&	0.90&	0.85&	0.81&	0.87&	0.75&	0.61\\
    \bottomrule
    \end{tabular*}

    \end{table*}
    \end{center}

\begin{figure*}[htbp]
\centerline{\includegraphics[width=1\linewidth]{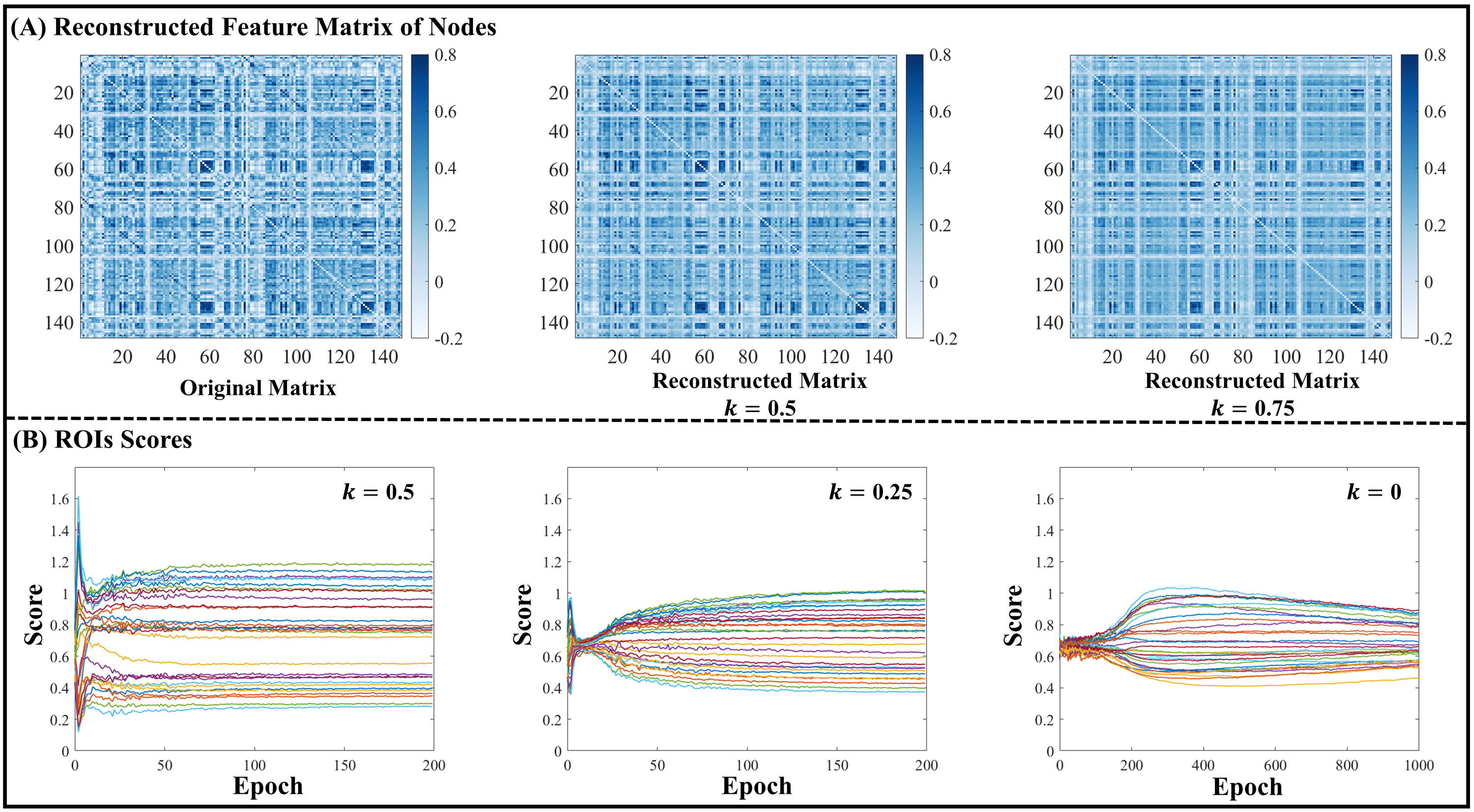}}
\caption{Effects of k on Reconstruction and ROI Scores. (A) Reconstruction matrix for a randomly selected individual in the SOCIAL task (S) at varying levels of k. (B) Variation in ROI scores over epochs at different k values for the REST task (RS).\label{fig4}}
\end{figure*}

\end{enumerate}

\subsection{Distribution of High-Score ROIs}

\begin{figure*}[htbp]
\centerline{\includegraphics[width=1\linewidth]{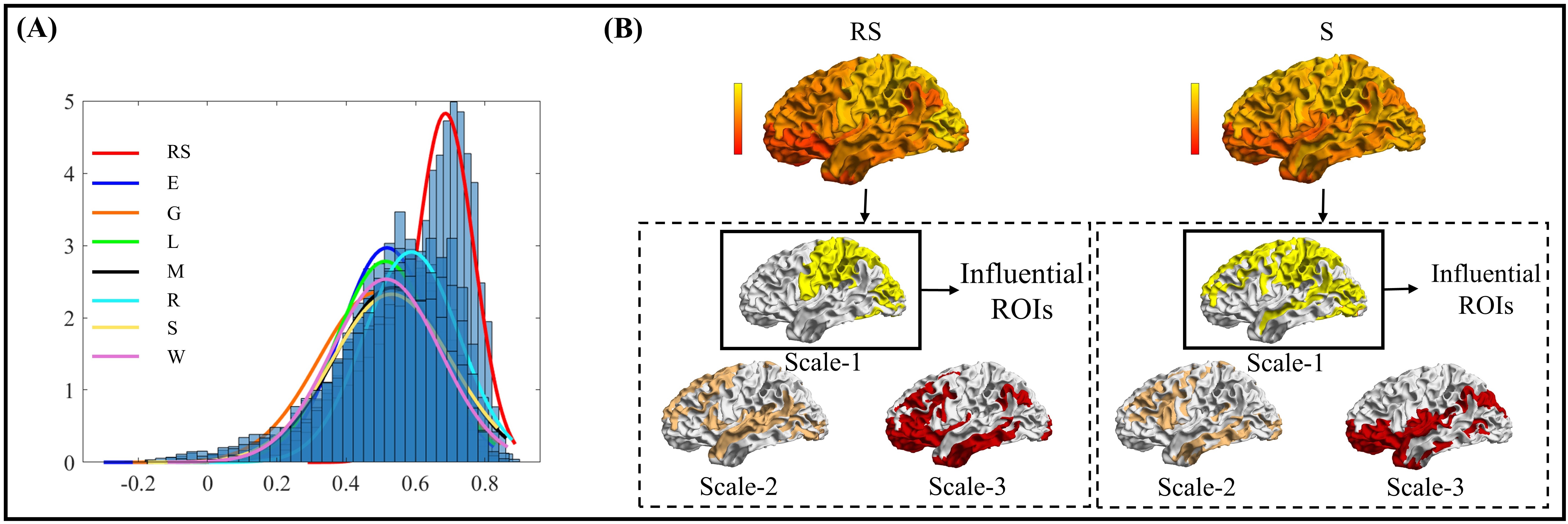}}
\caption{Group-level scores of ROIs. (A) represents the histogram of inter-individual ROIs scores similarity of eight sets of experiments. (B) represents the scores of all ROIs and the division of ROIs at RS and S.\label{fig5}}
\end{figure*}

Through the experiments in Section 3.1, for each of the eight sets of experiments, we obtain the contribution scores of brain ROIs of 98 individuals. According to Section 2.4, we need to obtain group-level scores. Through Pearson correlation analysis of inter-individual scores, we discover that scores of ROIs among individuals are similar for each set of experiment, as shown in \textbf{Fig. \ref{fig5} (A)}. That is, from a statistical point of view, the scores are very similar between individuals. Therefore, we can obtain a group-level score results by a simple method. By averaging the scores of the corresponding brain ROIs of all individuals in the same experiment, we obtain group-level scores. 

For the group-level scores of the brain ROIs at each set of experiments, ROIs are divided into three scales according to the scores from high to low. Each scale contains 20\%, 30\%, and 50\% of the ROIs, and the corresponding number of ROIs is 30, 44, and 74 respectively. Scale-1 ROIs are considered as I-nodes. Scale-2 and Scale-3 are in descending order of importance in brain networks. Using experiments RS and S as examples, the scales are visualized on brain surface, as shown in \textbf{Fig. 5 (B)}.

From \textbf{Fig. 5 (B)}, the distribution of the three scales is interesting. For RS, the Scale-1 is more distributed in the parietal and occipital parts, and the Scale-2 is distributed in the prefrontal, auditory, and visual parts. The Scale-3 is located at the bottom of the frontal and temporal lobes. For the task fMRI experiments, the distribution of the three scales is similar to RS, and the difference is mainly in the frontal lobe. For example, the Scale-1 of S is partially distributed in the prefrontal lobe in addition to the parietal and occipital parts.

\begin{figure*}[htbp]
\centerline{\includegraphics[width=0.8\linewidth]{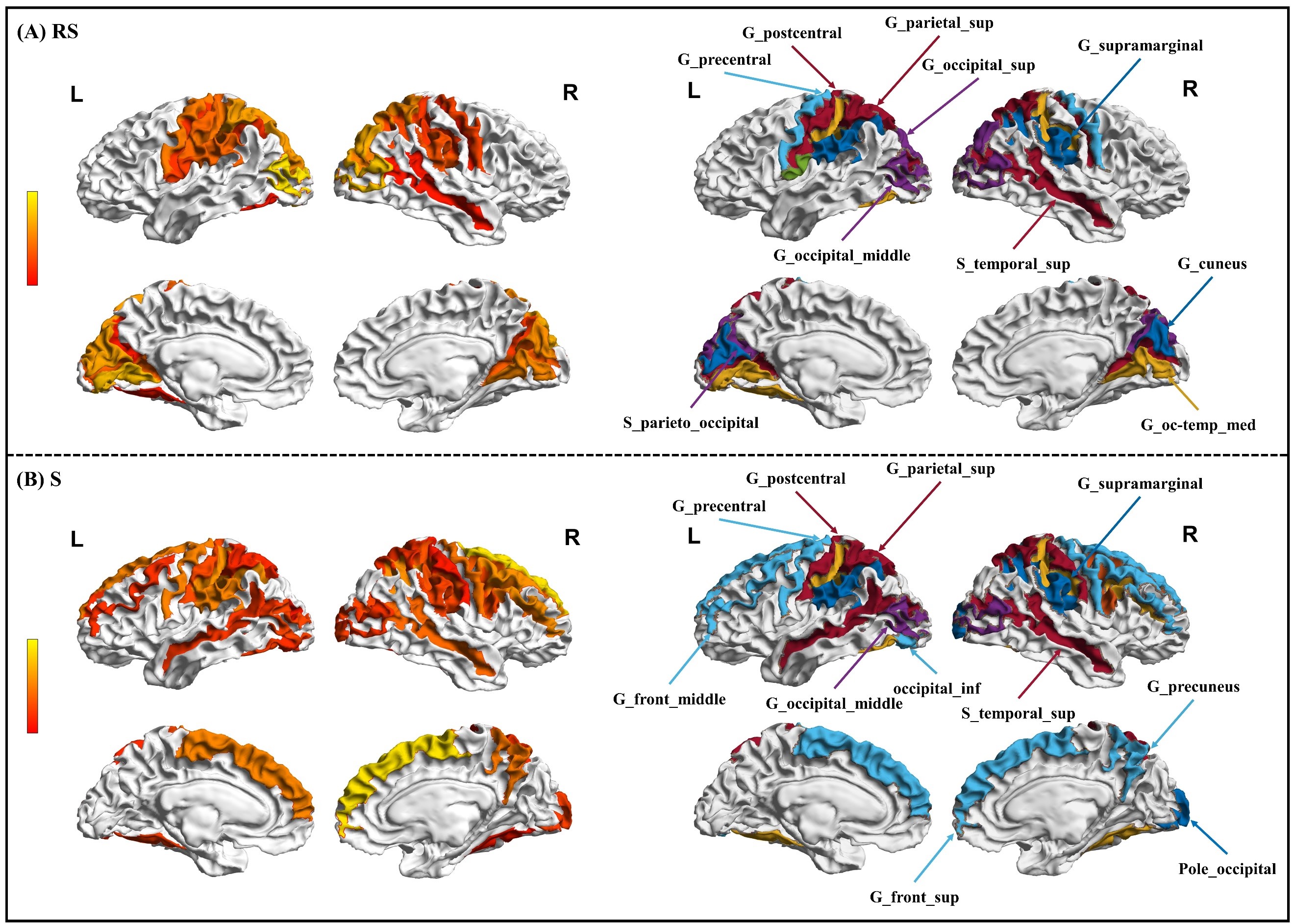}}
\caption{The Scale-1 ROIs in RS and S.\label{fig6}}
\end{figure*}

We focus more on Scale-1 ROIs and visualize them. Taking RS and S as examples, we show the brain ROIs involved in Scale-1, as well as the scores, as shown in \textbf{Fig. \ref{fig6}}. We have indicated the names of some of the ROIs, and some common important ROIs are present in both RS and S, such as the Superior parietal, precentral gyrus, postcentral gyrus, and occipital region. In addition, the results of all experiments are shown in the supplementary \textbf{Fig. \ref{fig1}}. For all sets of experiments, Scale-1 ROIs are distributed in the superior frontal lobe, lateral parietal lobe, and lateral occipital lobe of brain. Moreover, there are a certain number of replication ROIs in Scale-1 ROIs in the eight sets of experiments. And even if a ROI exists in Scale-1 ROIs of multiple sets of experiments at the same time, its score in different experiments is different, that is, its relative importance is different.

\subsection{High Hub Attributes of I-nodes}
\begin{figure*}[htbp]
\centerline{\includegraphics[width=1\linewidth]{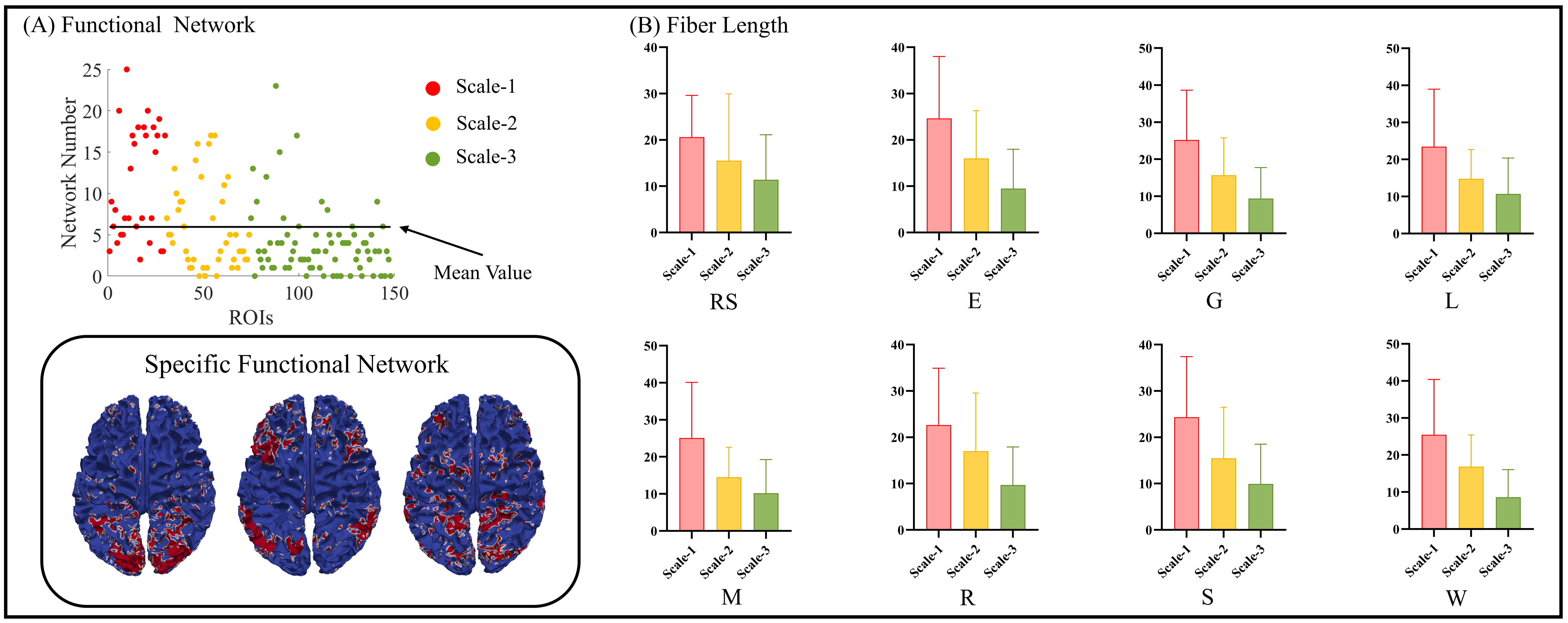}}
\caption{The Scale-1 ROIs in RS and S.\label{fig7}}
\end{figure*}

According to Section 2.5, we analyze the significance of reconstructing important nodes in the brain network. In terms of function, we take RS as an example to analyze the participation degree of high-score ROIs in the functional network. We conduct dictionary learning on functional signals of rs-fMRI and divide functional signals of the whole brain into multiple meaningful brain functional networks. And then count the number of brain ROIs in different Scales participating in the brain functional networks, as shown at the top of \textbf{Fig. \ref{fig7} (A)}. It shows that the brain high-score ROIs involve more brain networks than other ROIs, which means that the high-score ROIs have greater functional significance and play a key role in the functional integration of the brain. We list some brain functional networks involved in high-score ROIs, such as visual networks, as shown at the bottom of \textbf{Fig. \ref{fig7} (A)}. In terms of structure, we count the length of fibers passing through all brain ROIs, as \textbf{Fig. \ref{fig7} (B)}. It shows that in all sets of experiments, long fiber connections are more concentrated in the high-score ROIs rather than other ROIs, indicating that the high-scale ROIs occupy central positions in the structural connectivity and are critical for the propagation of information. This is similar to the pattern of connections found in brain hubs in previous studies \cite{nie2012axonal}.

\begin{figure*}[htbp]
\centerline{\includegraphics[width=1\linewidth]{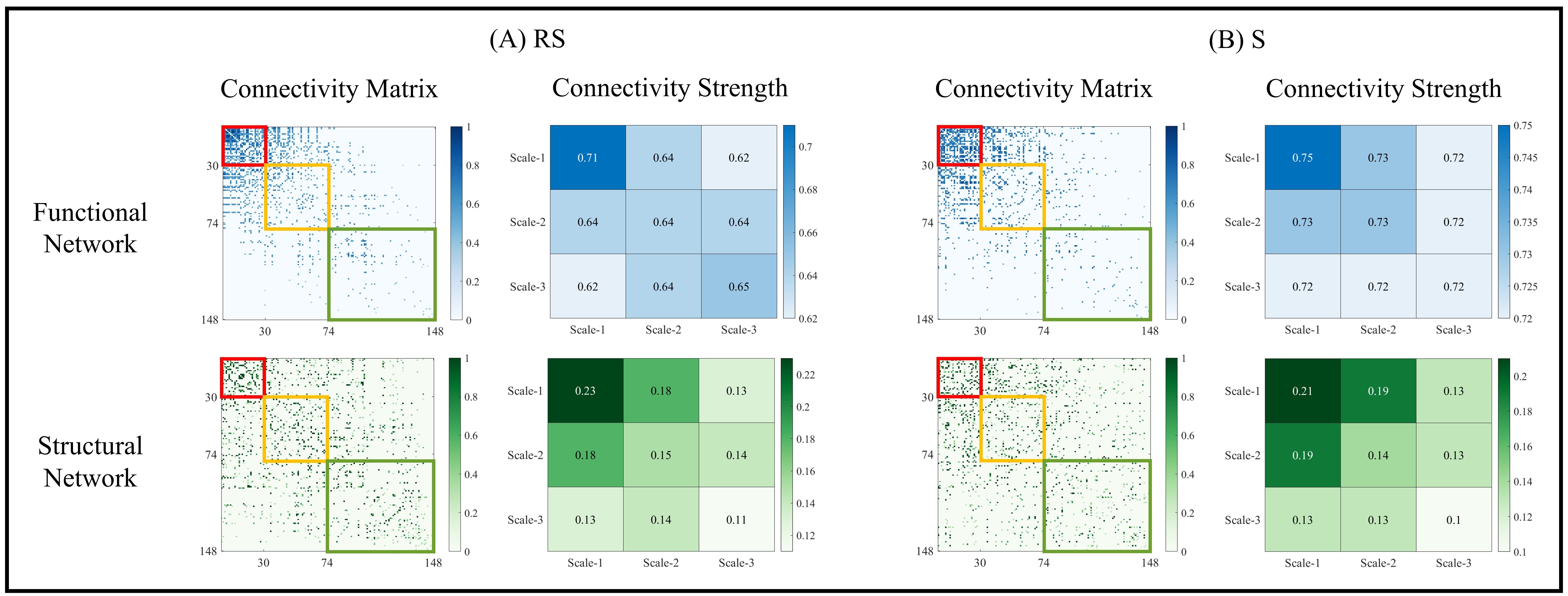}}
\caption{Network connectivity strength at RS and S. The red, yellow, and green frames represent connections within the Scale-1, Scale-2, and Scale-3 ROIs, respectively.\label{fig8}}
\end{figure*}

We also verify whether Scale-1 ROIs are important nodes in the brain network through graph theory attributes. First, we need to build brain functional and structural connectivity networks. For functional networks, we average the functional correlation matrix of all individuals under the same experiment, and then take a threshold value to retain 7\% edges. For the structural network, we average the structural connectivity matrix of all individuals, and then take a threshold value to retain 7\% edges, too. In the end, we have one structural connectivity network and eight functional connectivity networks. The network connections are visualized as a heat map, where ROIs are ranked from lowest to highest score, as shown in \textbf{Fig. \ref{fig8}}.

The results in \textbf{Fig. \ref{fig8}} show that for functional networks (blue), the distribution of connections varies from dense and strong to sparse and weak from top left to bottom right. This indicates that high-score ROIs have strong connections in the network, while low-score ROIs have weak connections in the network. Structural networks (green) have distributed representations similar to functional ones, but not as strong as functional networks. The functional and structural connectivity strength is relatively similar overall, with decreasing strength of connections in the Scale-1, Scale-2, and Scale-3. 

\begin{figure*}[htbp]
\centerline{\includegraphics[width=1\linewidth]{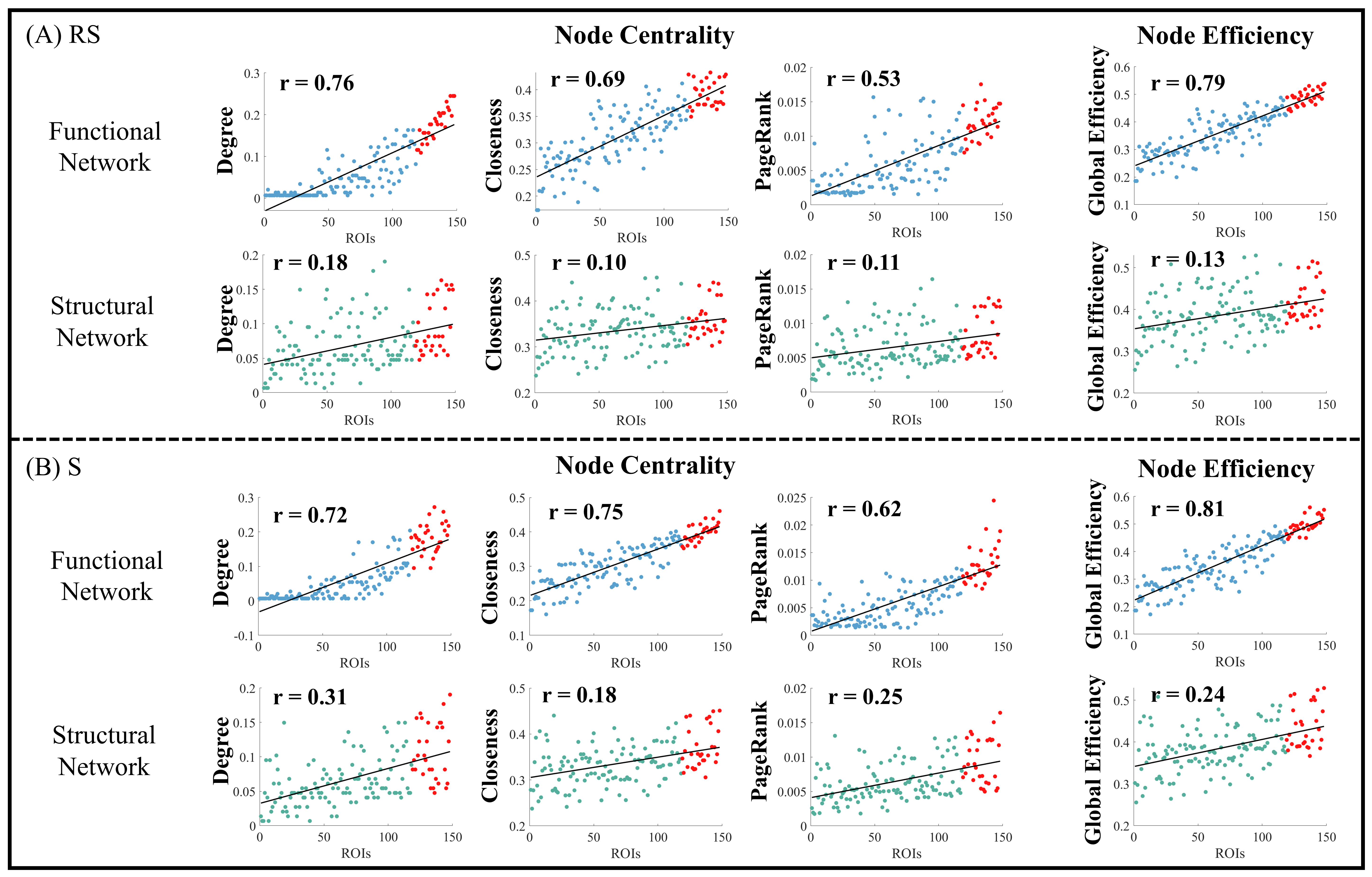}}
\caption{Graph metrics of ROIs with scores, including node centrality and node efficiency. The horizontal axis indicates the ROIs ranked from lowest to highest scores, and Red indicates the Scale-1 ROIs.\label{fig9}}
\end{figure*}

For the measure of network centrality and efficiency, we calculate the metrics of all ROIs in functional networks (blue) and structural networks (green) respectively, as shown in \textbf{Fig. \ref{fig9}}. We utilize three centrality metrics, namely degree centrality, closeness centrality and PageRank centrality. It shows that the metrics are positively correlated with scores of ROIs. We also analyze the global efficiency of ROIs, which is still positively correlated with scores of ROIs. In the functional network and structural network, high-score ROIs have high node efficiency. Therefore, we consider the Scale-1 ROIs as I-nodes, and the network they form assumes the main information transfer function in the functional and structural network.

\subsection{Comparison with Rich-Club }
\begin{figure*}[htbp]
\centerline{\includegraphics[width=0.76\linewidth]{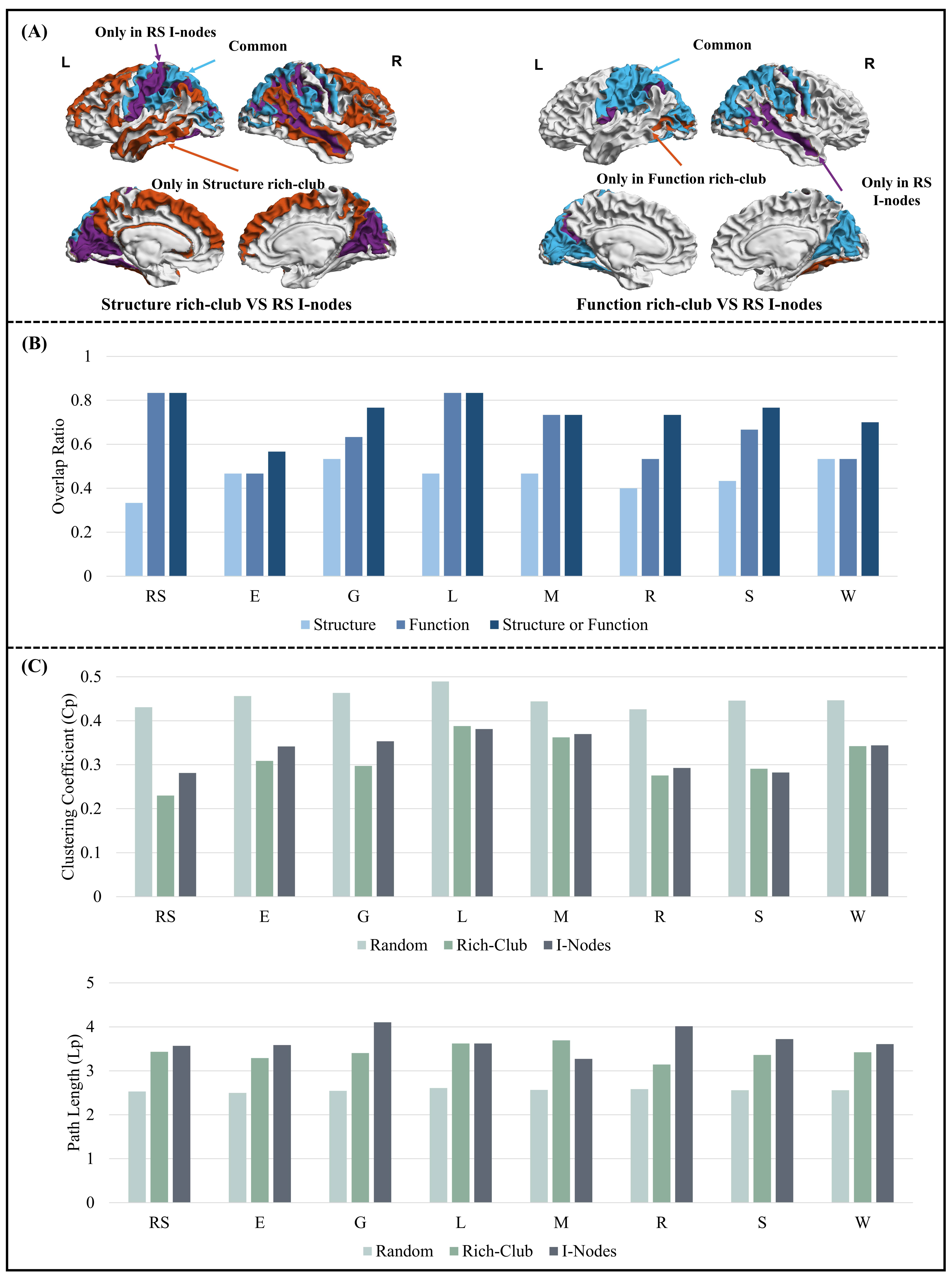}}
\caption{Comparison of our I-nodes with rich-club. (A) represent the comparison of RS I-nodes and structural rich-club, resting state functional rich-club on the brain surface. (B) represents the overlap ratio between I-nodes and rich-club in all experiments and (C) represent the comparison of the impacts of I-nodes and rich-club on network properties. The top and bottom of it represent the clustering coefficient and the characteristic path length of the network, respectively.\label{fig10}}
\end{figure*}

We compare the obtained brain I-nodes obtained by the proposed method with brain rich-club \cite{van2011rich}. The functional network and structural network of the brain obtained in Section 3.3 are binarized, and then $\varPhi (k)$ in different networks are calculated separately \cite{van2011rich}. The selection principle of the rich-club is (1) $\varPhi (k)>1$ and as high as possible; (2) The number of nodes in the rich-club is close to the number of I-nodes. Thus, we obtained the structure rich-club and eight sets of functional rich-clubs. We calculated the overlap ratio between I-nodes and rich-club, as shown in \textbf{Fig. \ref{fig10} (B)}. For each set of experiments, the three bars from left to right (light, medium, and dark blue) indicate the proportion of structural rich-club, functional rich-club, and structural or functional rich-club in our I-nodes, respectively. The proportion of overlap between the I-nodes and the structural rich-club is smaller than that between the functional rich-club, indicating that although our multimodal approach combines structure and function, it is more influenced by function. This may indicate that the I-nodes of the human brain are more composed of regions that represent brain function.

The proportion of structural rich-club or functional rich-club in the I-nodes is between 0.57 and 0.84, indicating that the I-nodes we obtained are partially consistent with rich-club, and reveal some new important brain regions. Taking RS as an example, we visualize the I-nodes and rich-club on the brain surface, as shown in \textbf{Fig. \ref{fig10} (A)}. The frontal (frontal\_sup, frontal\_middle), parietal (parietal\_sup, parietal\_inf-Supramar, parietal\_inf-Angular), occipital (occipital\_sup, occipital\_middle), temporal (temporal\_middle), and precuneus form the structural rich-club. While some overlap exists between the structural and the functional rich-club, the latter is more concentrated in the occipital (occipito-temporal gyrus) and central ROIs (postcentral gyrus). It is worth noting that the precuneus and frontal ROIs are not in the functional rich-club, and they are also absent from our RS I-nodes (the top 20\%, the precuneus (R) ranking 35, 24\% in RS I-nodes), perhaps related to the fact that multimodal results are more functionally influenced. Most of our RS I-nodes coincide with resting state functional rich-club, and some new ROIs are revealed, which are mainly concentrated in the parieto-occipital sulcus and central (subcentral) ROIs. 

In addition, in order to compare the importance of rich-club and I-nodes to the brain network, we removed rich-club and I-nodes respectively, and compared the global properties of the resulting 2-group networks, as shown in \textbf{Fig. \ref{fig10} (C)} We also added a control group, that is, the network obtained by removing random regions. The three bars in \textbf{Fig. \ref{fig10} (C)} from left to right (light, medium, and dark green) show the elimination of random regions, rich-club, and I-nodes, respectively. Compared with the random regions, the elimination of rich-club and I-nodes results in lower clustering coefficients and higher characteristic path lengths. This suggests that the elimination of rich-club and I-nodes has more serious damage to the brain network. For the elimination of rich-club and I-nodes respectively, the elimination of rich-club shows low clustering coefficient and characteristic feature path length, while the elimination of I-nodes shows high clustering coefficient and high characteristic path length. This shows that the two have different effects on the brain network. rich-club may favor the local features of the network, while I-nodes may favor the global features of the network.

\subsection{The Foundational Elements Coordinated Structure and Multi-tasks}
\begin{figure*}[ht]
\centerline{\includegraphics[width=0.8\linewidth]{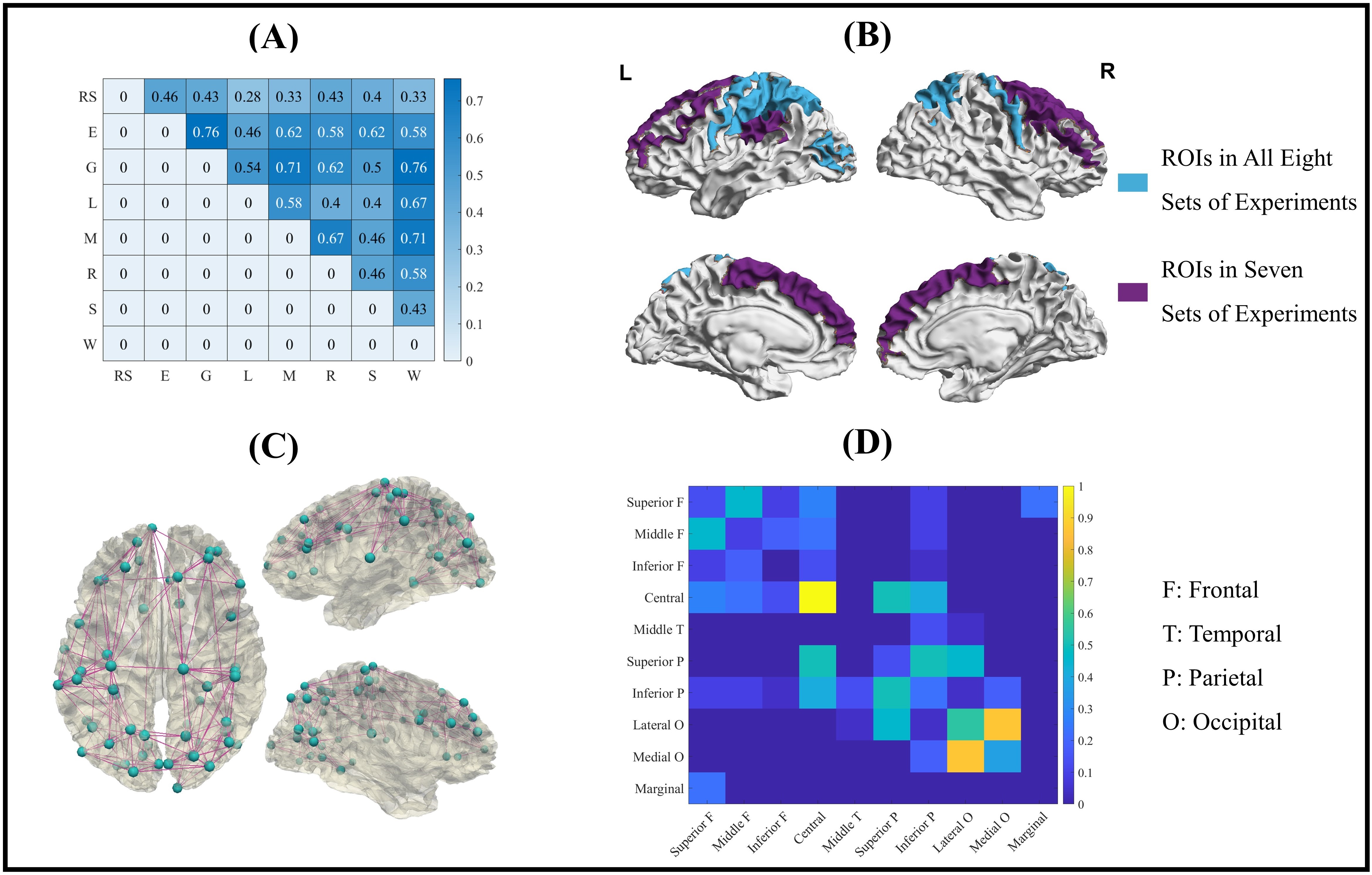}}
\caption{Joint analysis of I-nodes for different functions. (A) represents the IoU of Scale-1 ROIs between experiments. (B) represents the ROIs exits in most of experiments. (C) and (D) represent common connections and distributions, respectively.\label{fig11}}
\end{figure*}

As shown in Section 3.2, there is a certain overlap among the Scale-1 ROIs in the eight sets of experiments. We list the specific ROIs of Scale-1 of all experiments, which involved 56 ROIs in total, as shown in supplementary \textbf{Fig. \ref{fig2}}. We visualize some of these ROIs of the brain, as shown in \textbf{Fig. \ref{fig11}}. Blue represents the ROIs where all experiments are activated, purple represents the ROIs where seven sets of experiments are activated. It can be observed that the superior frontal, central sulcus, superior parietal, and middle occipital are activated in most sets of experiments, indicating that they play a very important role in both the rs-fMRI and task fMRI. In addition, we also calculate the Intersection over Union (IoU) of Scale-1 in different experiments, as shown in \textbf{Fig. \ref{fig11} (A)}. The value between E and G is the highest, indicating that the functional activation of the two tasks is similar.

Based on the 56 ROIs that included Scale-1 ROIs of all experiments, we obtain common connections for the structure and multi-functions, as shown in \textbf{Fig. \ref{fig11} (C)}. Specifically, the functional connections of each task in the common ROIs are counted first, then the connections that occur more than 5 times in all tasks are counted, and finally, the intersection of the structural connections is taken. It shows that these connections are more evenly distributed throughout the brain, rather than being concentrated in some particular regions. It can be concluded from \textbf{Fig. \ref{fig11} (D)} that the intra-occipital and intra-central connections are the most numerous, while the frontal and parietal connections are more evenly distributed. This indicates that the occipital region plays an important role in the common connections. The I-nodes we have identified are useful for exploring important regions that coordinate brain structure and various functions, providing new insights into how brain structure and function are linked together.

In addition, we analyze the implications of the I-nodes in terms of cognitive neuroscience. The I-nodes are decoded by meta-analysis and the terms related to mental function are obtained. As shown in \textbf{Fig. \ref{fig11} (B)}, both blue and purple ROIs exist in most of experiments, which called Common nodes. In each group of experiments, the I-nodes outside the Common nodes are called Special nodes. The Special nodes corresponding to experiment RS, E, G, L, M, R, S and W are called as Spe\_RS, Spe\_E, Spe\_G, Spe\_L, Spe\_M, Spe\_R, Spe\_S, and Spe\_W, respectively. For the Common nodes and these eight groups of Special nodes, we adopt BrainStat \cite{lariviere2023brainstat}to decode their mental function and get the distribution of terms related to them. For each group of nodes, we take the six terms most relevant to it for analysis, as shown in \textbf{Table \ref{tab3}}.

    \begin{table}[!ht]%
    \begin{center}
    \caption{The six terms of mental function most relevant to each group of nodes by meta-analysis\label{tab3}}%
    \begin{tabular*}{\textwidth}{@{\extracolsep\fill}ccccccc@{\extracolsep\fill}}%
    \toprule
    \textbf{I-nodes} & \multicolumn{6}{c}{\textbf{Terms of mental function}}  \\
    \midrule
\textbf{Common}&	Intention&	Demand&	Motor control&	Reasoning&	Movement&	Working memory\\
\textbf{Spe\_R}S&	Face recognition&	Object recognition&	Visual attention&	Spatial attention&	Mental imagery&	Selective attention\\
\textbf{Spe\_E}	&Spatial attention&	Visual	&Mental imagery&	Object recognition&	Imagery	&Categorization\\
\textbf{Spe\_G}&	Spatial attention	&Verbal fluency&	Visual	&Selective attention&	Judgment&	Reasoning\\
\textbf{Spe\_}L&	Decision making&	Intelligence&	Reasoning&	Memory retrieval&	Valence&	Verbal fluency\\
\textbf{Spe\_M}&	Verbal fluency&	Visual	&Fixation&	Spatial attention&	Decision making&	Judgment\\
\textbf{Spe\_R}&	Spatial attention&	Fixation&	Selective attention&	Object recognition&	Working memory	&Reasoning\\
\textbf{Spe\_S}&	Object recognition&	Spatial attention&	Visual attention&	Selective attention	&Face recognition&	Mental imagery\\
\textbf{Spe\_W}&	Reasoning	&Judgment&	Spatial attention&	Episodic memory	&Object recognition&	Memory retrieval\\

    \bottomrule
    \end{tabular*}
    \end{center}
    \end{table}

Our findings are shown as follow. First, the mental functions of Common nodes are quite different from that of the other eight groups. Common nodes involve a lot of motor-related terms, such as motor control and movement. And it is related to intention, reasoning, and working memory, which may be the basis of other functional tasks. Second, for the eight sets of Special nodes, there are terms such as visual, spatial attention, selective attention that appear several times, which are related to task execution. For some tasks, we get terms that are closely related to it, such as judgment in Spe\_G, verbal fluency in Spe\_L, and episodic memory in Spe\_W. Such findings suggest that novel insights can be detected by exploring the brain I-nodes. Due to the strong correlation between the I-nodes and brain functions, it has important application potential in the field of brain analysis. On one hand, it is very possible to better predict individual behavior based on I-nodes rather than whole brain. On the other hand, I-nodes can serve as potential biomarkers for neurological diseases.

\subsection{Stability of I-nodes on New Brain Atlas}
\begin{figure*}[ht]
\centerline{\includegraphics[width=0.8\linewidth]{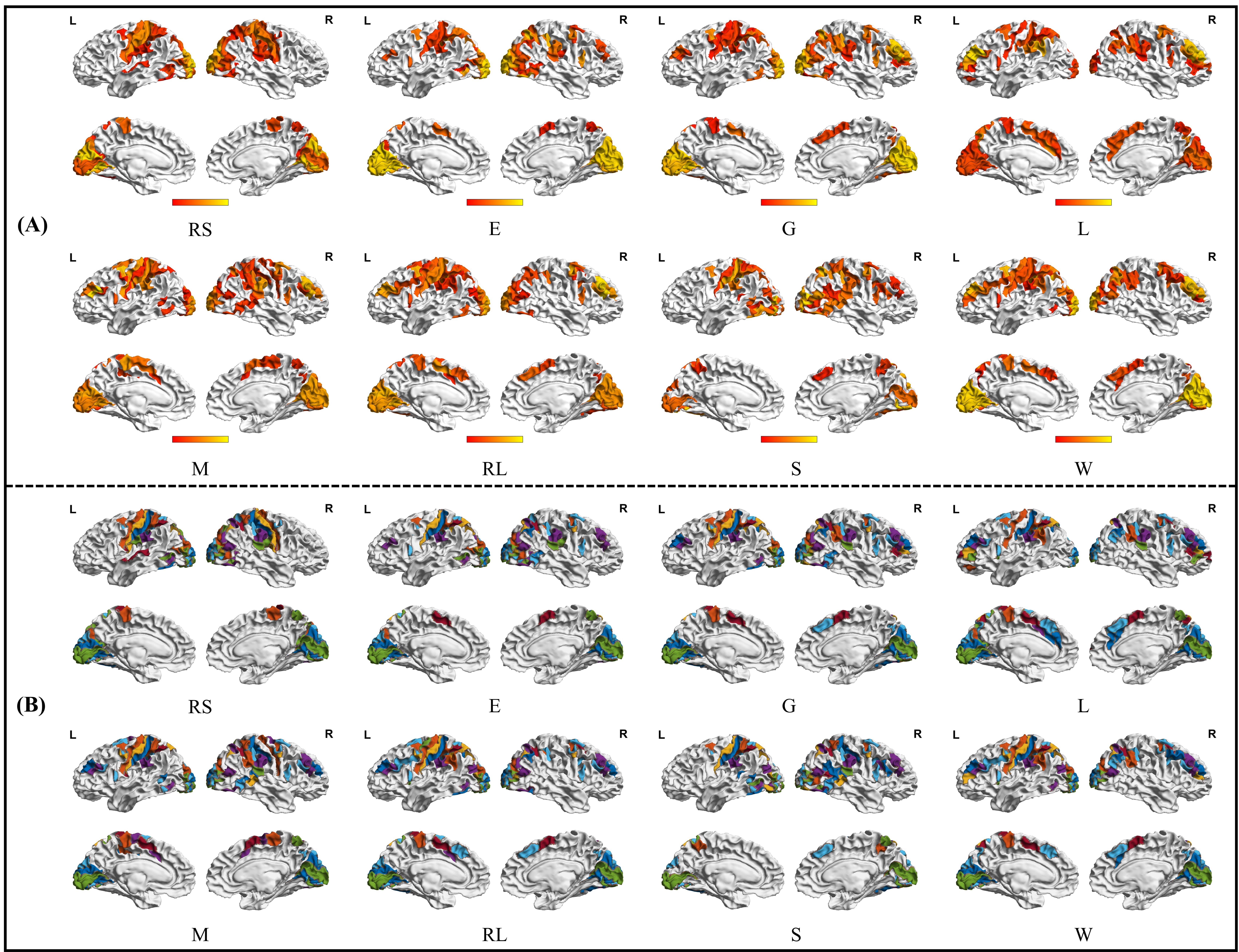}}
\caption{The Scale-1 ROIs in all experiments. (A) and (B) represent scores of ROIs and boundaries of ROIs, respectively.\label{fig12}}
\end{figure*}

In order to verify the stability of our proposed model, we use a new atlas on the original data set for experiments. As same as Atlas\_1, we still select the top 20\% ROIs with high scores as Scale-1, as shown in \textbf{Fig. \ref{fig12}}. Overall, the Scale-1 ROIs for Atlas\_1 and Atlas\_2 are similar. For example, Scale-1 ROIs for Atlas\_2 are also concentrated in the central gyrus, superior parietal, and occipital lobes. Unlike the Scale-1 for Atlas\_1, the Scale-1 for Atlas\_2 is not as widely distributed in the frontal lobe, which is also related to the more finely division of Atlas\_2.

\begin{figure*}[ht]
\centerline{\includegraphics[width=1\linewidth]{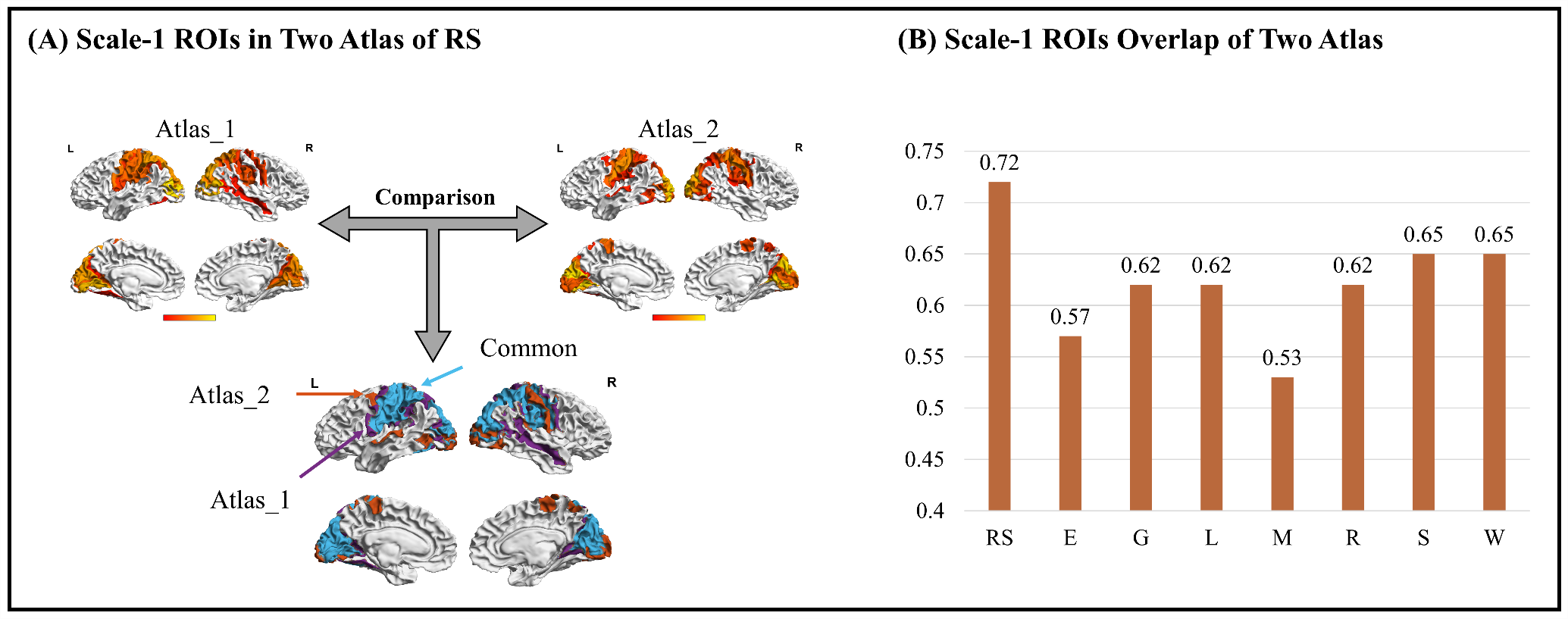}}
\caption{Overlap of Scale-1 ROIs between Atlas\_1 and Atlas\_2. (A) represents the visualization of Scale-1 ROIs of the two atlases on the brain in RS. (B) represents the overlap of the Scale-1 ROIs of the two atlases in all sets of experiments.\label{fig13}}
\end{figure*}

In addition, we calculate the overlap of the Scale-1 ROIs of the two atlases, as shown in Fig. 13. In \textbf{Fig. \ref{fig13} (A)}, the Scale-1 ROIs of the two atlases in the RS experiment and their overlap are visualized. Blue represents ROIs that exist in both atlases, and purple and orange represent ROIs that exist only in Atlas\_1 or only in Atlas\_ 2, respectively. Blue occupies most of the ROIs. Orange and purple ROIs also exist around the blue ROIs. There are two reasons for the differences in results. First, the ROI divisions in the two atlases do not completely overlap. Second, the different atlases themselves introduce data variations that lead to divergent results.

The Scale-1 ROIs overlapping degree of the two atlases in all experiments is shown in Fig.13 (\textbf{Fig. \ref{fig13} (B)}). The specific process of calculation is as follows: we count the Scale-1 ROIs’ area (here we use the number of brain vertices) S1 and S2 of the two atlases respectively, and then divide the area S of the overlapping area by S2 to get the result. The overlapping degree of all experiments is about 0.6, among which RS had the highest coincidence degree, and E and M had the lowest overlapping degree. The high degree of overlap of RS may also reflect the stable characteristics of the RS signal itself, which is less affected by the atlas division. The experimental results show that the proposed model is stable and can be used as a reproducible method to explore the I-nodes of other data. 

\section{DISCUSSION AND CONCLUSION}\label{sec4}

We propose a self-supervised graph reconstruct framework based on Graph-Transformer to identify I-nodes by fusing multimodal neuroimaging data. It includes: (1) Through self-supervised modeling, the regions that contribute the most to reconstructing the brain graph are considered influential; (2) The model established on the Graph-Transformer can extract features well from the brain graph by combining local and global connectivity information; and (3) Fusing graph-based brain structure and function data provides a comprehensive understanding of how the brain network architecture arises from multimodal connections. We conduct separate experiments to integrate the structure and different functional data, and obtained the I-nodes under different integration conditions, and there are some common regions among them.

The I-nodes we obtained have an important place in the brain network and largely overlap with previously identified important regions, such as the rich-club. For example, in RS integrating structure and rs-fMRI, the located I-nodes mainly include: sensorimotor cortex (such as precentral gyrus and postcentral gyrus), visual cortex (such as cuneus, middle occipital gyrus, and superior occipital gyrus), superior parietal lobule, superior temporal lobe, and supramarginal gyrus. Among them, the I-nodes of the visual cortex, superior parietal lobule, and superior temporal lobe correspond to previous studies on brain hubs. Sporns et al. summarized that the superior temporal gyrus, superior parietal lobule, and occipital regions are the central locations of the structural network, and the superior parietal lobule is the hub of the resting state functional network\cite{van2013network}. Achard et al. defined the parietal and occipital regions as functional hubs of the resting state network through average path length \cite{achard2006resilient}. Tomasi et al. discovered high levels of local functional connectivity density (IFCD) in the occipital cortex such as the cuneus, so it is considered a prominent functional hub in the brain \cite{tomasi2010functional}.

However, we also observed some differences. The sensorimotor cortex and the supramarginal gyrus are identified as I-nodes, which differed from some previous works on brain hubs. Some studies suggest that although the sensorimotor cortex plays a key role in the somatosensory and motor control networks, it is not a hub node in the traditional sense of the brain network \cite{tomasi2011association}. Through the discussion in Section 3.4, we believe that the I-nodes are more affected by brain function, and the sensorimotor cortex has a high degree of centrality in the resting state functional network, which is the influential regions. This is consistent with the findings of Battiston et al. \cite{battiston2018multiplex} in their exploration of multimodal cores. And the role of the sensorimotor cortex in the joint network of structural and resting state functional networks should be emphasized. The supramarginal gyrus has been shown to mediate information transmission between multiple advanced cognitive networks\cite{carter2013nexus} , making it a potential influential region in the brain. 

In addition, some common I-nodes exist in the results of integrating structure and other different functions. They are respectively the frontal gyrus, superior parietal lobule, middle occipital gyrus, supramarginal gyrus, and sensorimotor cortex. We believe that these regions play an important role in coordinating structure and different functions, and they may work closely together to complete the integration of information in advanced cognitive tasks. For example, Cole et al. confirmed that the lateral prefrontal cortex and the right premotor cortex are the core of cognitive control \cite{cole2012global}

Overall, this paper proposes a new insight to detect the I-nodes of the brain network by the self-supervised method based on reconstruction. First, Through the proposed model, the brain nodes that are important for the reconstruction task are obtained and considered as the I-nodes. Second, we analyze the functional and structural properties of the I-nodes by means of functional networks and fiber connections. Through the graph theory attribute, it is proved that the I-nodes have the hub attribute. And our method is compared with the rich-club. Finally, in order to verify the stability of I-nodes extracted from the model, we test the model on the new atlas and obtained good stability. In this work, we explore a novel approach to explore I-nodes using self-supervised learning, which provides a new track to explain how the brain works. In future work, we will conduct in-depth studies on the working mechanism of the brain according to I-nodes. For example, studying the hierarchical or dynamic characteristics of brain networks based on the graph of I-nodes. Besides, I-nodes can be treated as the biomarkers for the brain disease diagnosis.

\section{ACKNOWLEDGMENT}\label{sec5}

This work was supported by National Natural Science Foundation of China (62376219 and 62006194); Foundational Research Project in Specialized Discipline (Grant No. G2024WD0146); Faculty Construction Project (Grant No. 24GH0201148)

\bibliography{wileyNJD-AMA}

\end{document}